\documentclass[letter,fleqn,usenatbib]{mnras}
\pdfoutput=1

\usepackage{txfonts}   

\usepackage[T1]{fontenc}
\usepackage{ae,aecompl}

\usepackage{graphicx}	
\usepackage{amssymb}	

\newcommand{\bpp}{BP\,Psc }
\newcommand{\bppb}{BP\,Psc}
\newcommand{\ff}{f\mbox{}f}
\newcommand{\srcom}{\rm}
\newcommand{\ercom}{\rm}

\hypersetup{pdfauthor={J. de Boer},
               pdftitle={BP Piscium: its flaring disk imaged with SPHERE/ZIMPOL},
               pdfkeywords={circumstellar matter, protoplanetary discs, stars: evolution, techniques: polarimetric, polarization, techniques: high angular resolution},
               bookmarksnumbered=true}

\title[BP\,Piscium: imaging of its disk with SPHERE/ZIMPOL]
{BP\,Piscium: its f\mbox{}laring disk imaged with SPHERE/ZIMPOL\thanks{
The observations were taken during SPHERE science verif\mbox{}ication by the European Southern Observatory, Chile (ESO program ID: 60.A-9375(A)}}

\author[J.~de Boer et al.]
{
J.~de Boer,$^{1,2}$\thanks{E-mail: deboer@strw.leidenuniv.nl}
J.\,H. Girard,$^{2}$
H.~Canovas,$^{3}$ 
M.~Min,$^{4,5}$
M.~Sitko,$^{6,7}$
C.~Ginski,$^{1}$
\newauthor
S.\,V.~Je\ff ers,$^{8}$
D.~Mawet,$^{9}$
J.~Milli,$^{2}$
M.~Rodenhuis,$^{1}$
F.~Snik$^{1}$ and
C.\,U.~Keller$^{1}$.
\vspace{2mm}
\\
$^{1}$Leiden Observatory, Universiteit Leiden, P.O. Box 9513, 2300 RA Leiden, The Netherlands.\\
$^{2}$European Southern Observatory, Casilla 19001, Santiago, Chile.\\
$^{3}$Departamento de F\'isica Te\'orica, Universidad Aut\'onoma de Madrid, Cantoblanco 28049 Madrid, Spain.\\
$^{4}$SRON Netherlands Institute for Space Research, Sorbonnelaan 2, 3584 CA Utrecht, The Netherlands.\\
$^{5}$Astronomical institute Anton Pannekoek, University of Amsterdam, Science Park 904, 1098 XH, Amsterdam, The Netherlands.\\
$^{6}$Department of Physics, University of Cincinnati, Cincinnati, OH 45221-0011, USA\\
$^{7}$Center for Extrasolar Planetary Studies, Space Science Institute, Boulder, CO 80301, USA\\
$^{8}$Institut fuer Astrophysik, Georg-August-Universitaet Goettingen, Friedrich-Hund-Platz 1, D-37077 Goettingen, Germany\\
$^{9}$Department of Astronomy, California Institute of Technology, 1200 E. California Blvd., Pasadena, CA, 91125, USA
}

\date{Accepted 2016 October 20. Received 2016 October 20; in original form 2016 May 15}

\pubyear{2016}

\begin{document}
\label{firstpage}
\pagerange{\pageref{firstpage}--\pageref{lastpage}}
\maketitle

\begin{abstract}
   Whether BP Piscium (\bppb) is either a pre-main sequence T\,Tauri star  at $d \approx 80$\,pc,
   or a post-main sequence G giant at $d \approx 300$\,pc is still not clear.
   As a f\mbox{}irst-ascent giant, it is the f\mbox{}irst to be observed with a molecular and dust disk.
   Alternatively, \bpp would be among the nearest T\,Tauri stars with a protoplanetary disk (PPD).
   We investigate \srcom whether \ercom the disk geometry resembles typical PPDs, 
   by comparing polarimetric images with radiative transfer models. 
   Our VLT/SPHERE/ZIMPOL observations allow us to perform Polarimetric Di\ff erential Imaging; 
   Reference Star Di\ff erential Imaging; and Richardson-Lucy deconvolution.
   We present the f\mbox{}irst visible light polarization and intensity images of the disk of \bppb.
   Our deconvolution conf\mbox{}irms the disk shape as detected before, 
   mainly showing the southern side of the disk.
   In polarized intensity the disk is imaged at larger detail and also shows the northern side,
   giving it the typical shape of high inclination f\mbox{}lared disks.
   We explain the observed disk features by
   retrieving the large-scale geometry with MCMax radiative transfer modeling, which yields
    a strongly f\mbox{}lared model, atypical for disks of T\,Tauri stars.
\end{abstract}



\section{Introduction}

\begin{figure*}
   \begin{center}
   \includegraphics[width=1\textwidth,trim = 0 15 0 15]{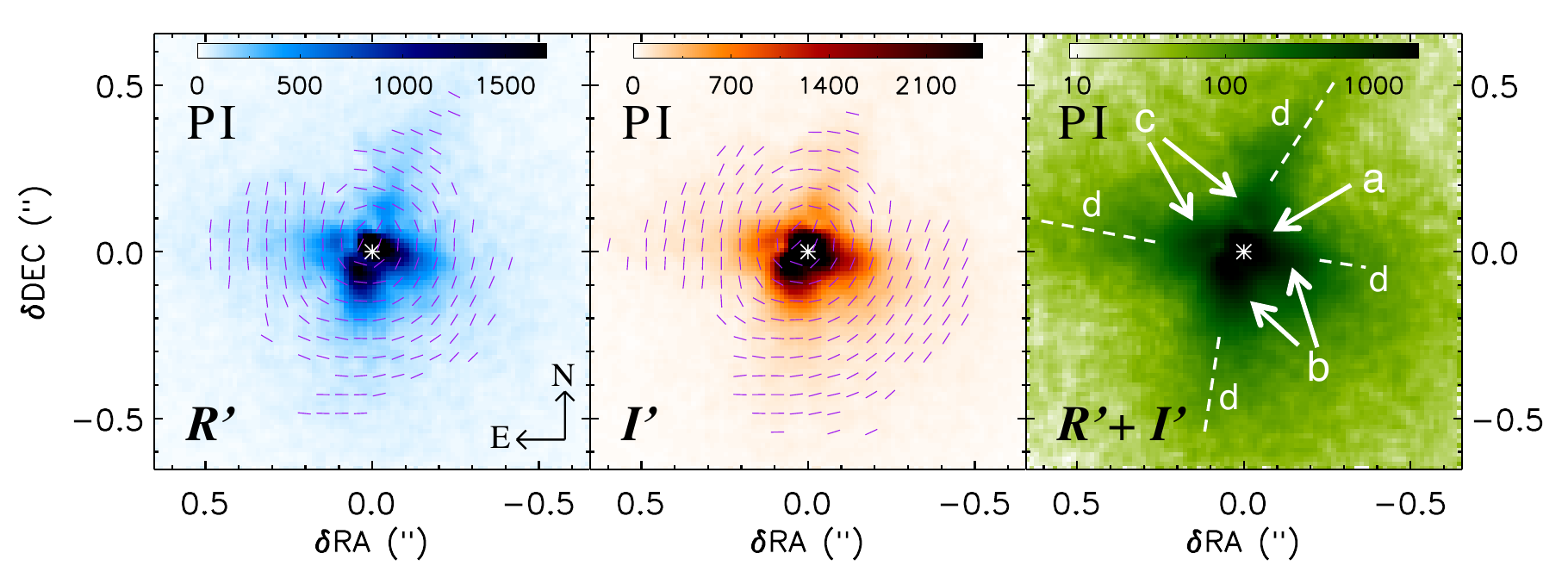}
   \end{center}
   \caption{ \label{fig:ipc}
	Polarized Intensity ($PI$) image of \bpp in $R'$-band (\textit{Left});
	in $I'$-band (\textit{Center}); and both f\mbox{}ilters combined(\textit{Right}).
	Where the signal is $> 10\times$ the background noise, purple lines show the polarization angle $P_\theta$.
	Note that the units are arbitrary and $R' +I'$ is presented in log scale. 
	The three panels all show the same features: 
	(\textbf{a}) a bright blob at position of the star center (highlighted with a white asterisk);
	(\textbf{b} and \textbf{c}) are the southern and northern `bowls' of the highly inclined disk;
	(\textbf{d}) extended f\mbox{}ingers, aligned with both the brighter bowls of the northern and southern disk.
      } 
\end{figure*}

Circumstellar disks appear at di\ff erent stages of stellar evolution.
During early stages of star formation, collapsing molecular clouds result in protoplanetary disks (PPDs) and later debris disks \citep{Williams:2011ARA&A}.
Although mass ejections of post-main sequence stars are often spherical, 
they are known to create disk-like surroundings for (post)AGB stars \citep{Skinner:1998MNRAS, Jeffers:2014A&A, Kervella:2015A&A}.

\citet[Z08]{2008ApJ...683.1085Z} detect a circumstellar gas disk around BP Piscium (hereafter \bppb) 
in\,$^{12}$CO\,(3-2) with SMA and in\,$^{12}$CO\,(2-1) with OVRO interferometric measurements,
and by deconvolving Keck $H$ and $K'$ band images they detected a dust disk at high inclination ($i = 75 \pm 10^\circ$, 
with $i = 0^\circ$ for a face-on disk) and 
position angle \textit{PA}\,$ = 118 \pm 5^\circ$.
Since no reliable parallax has been determined, the distance ($d$) to the star is highly uncertain.
For two possible evolutionary scenarios with an e\ff ective temperature $T_\mathrm{e\ff} \sim 5000$\,K,
the luminosities are matched to observations by varying $d$.
Z08 propose two possible evolutionary scenarios for \bppb:
1) at a distance $d \approx 80$\,pc, it is one of the nearest pre-main sequence Classical 
T\,Tauri Stars (CTTSs) with an age $\approx 10$\,Myr; or
2) at $d \approx 300$\,pc, \bpp is a post-main sequence star of a few Gyr at its f\mbox{}irst-ascent, or
hydrogen shell burning phase.
%
A more luminous (therefore $d \sim 5000$\,pc) (post-) Asymptotic Giant Branch (AGB, helium shell burning) star is ruled out due to its large Tycho 2 proper motion \citep[$\Delta RA = 44.4 \pm 4.1$\,mas yr$^{-1}$, $\Delta Dec = -26.3 \pm 4.3$\,mas yr$^{-1}$,][]{Hog:2000A&A}.
For a star on the f\mbox{}irst-ascent giant branch, the associated molecular disk, accretion and Herbig Haro objects would be the f\mbox{}irst ever to be detected. 
The growing primary star would have recently enveloped a previous companion, hence
creating the disk.
Z08 favor the G giant scenario for \bpp, mainly because of low lithium abundance, low surface gravity,
and lack of an associated star forming region. 
The stellar photosphere, obscured by the disk in visible and near-infrared wavelengths, is 
directly detected in X-rays with CHANDRA \citep{Kastner:2010}. 
The authors argue that the ratio of X-ray over bolometric luminosity is
too low for typical CTTSs but does agree with other rapidly rotating G giant stars. 
Furthermore, the G giant scenario is supported by the disk modeling of \citet[M10]{2010ApJ...724..470M}: 
the disk model for the Herbig Be star HD 100546 matches the Spectral Energy Distribution (SED) of \bppb,
yielding an inner disk morphology which suggests that \bpp is too luminous to be CTTS
and leads them to suggest the presence of a massive planet carving a gap in the disk.
However, early studies of PPDs \citep[e.g.][]{Espaillat:2011ApJ, Andrews:2011ApJ, Dong:2012ApJ} have shown 
disk models based on the SED to be highly degenerate for many disk parameters, 
which can be solved by including high spatial resolution images of the disks in the analysis.

We compare visible light polarimetric images of \bppb's disk with radiative transfer modeling
to constrain the 3D geometry of the system.
The recent increase of detected and modeled PPDs will allow a 
qualitative assessment of the CTTS scenario.

\section{Observations and data reduction}
\subsection{Observations with VLT/SPHERE/ZIMPOL}
\label{sec:obs}
We observed \bpp and reference star TYC\,5259-446-1 (T52) 
on 6 December 2014, during science verif\mbox{}ication of the 
Spectro-Polarimetric High-contrast Exoplanet REsearch (SPHERE) instrument \citep{Beuzit:2008SPIE}, 
the new high-contrast imager of the Very Large Telescope (VLT).
We used the Zurich IMaging Polarimeter (ZIMPOL, \citet{Thalmann:2008SPIE} in f\mbox{}ield-tracking, polarimetric (P2) mode. 
We used the $R'$ ($\lambda_0 = 626.3$\,nm; $\Delta \lambda = 148.6$\,nm) and 
$I'$ ($\lambda_0 = 789.7$\,nm; $\Delta \lambda = 152.7$\,nm) f\mbox{}ilters simultaneously.
The f\mbox{}ilter choice implied the use of the `grey' beam splitter, which sends $20\%$ of all visible light to the 
SPHERE eXtreme Adaptive Optics \citep[SAXO,][]{Fusco:2014eo} wave front sensor (WFS)  
and the remaining $80\%$ to ZIMPOL. 
The stellar magnitudes and air mass $a$ during observation of \bpp  (V = 11.9\,mag, $a = 1.45 \pm 0.13$) and T52 (V = 11.4\,mag, $a = 1.14 \pm 0.02$) 
posed an additional challenge for the WFS. 
As a result, we reached a Strehl ratio in the $R'$-band of $S_{R'} \sim 2.8\%$ and a Full Width at Half Maximum (FWHM$_{R'}) = 50$\,mas.
In the $I'$-band, we reached $S_{I'} \sim 6.8\%$, FWHM$_{I'} = 40$\,mas.

We brief\mbox{}ly summarize how ZIMPOL works, but refer to 
\citet{2012SPIE.8446E..8YS} for details on the instrument.
A full polarimetric cycle consists of 4 exposures of $60$\,s for each of the 4 half wave plate (HWP) angles: 
$\theta_\mathrm{hwp} = 0^\circ$; $45^\circ$; $22.5^\circ$; $67.5^\circ$.
We recorded 1 polarimetric cycle ($=16$\,min on target) for T52 and 3 polarimetric cycles ($= 48$\,min on target) for \bppb.
During each individual exposure, the Ferro-electric Liquid Crystal (FLC) reverses the polarization state, 
by switching its fast axis from the initial (\textit{A}) state at $0^\circ$ to $45^\circ$ (\textit{B} state),
at a frequency of $26.97$\,Hz (= $1618 \times$ per integration). 
In the f\mbox{}irst frame (the \textit{0}-phase), `charge shu\ff \mbox{}ling' ensures that the light observed during the FLC's
\textit{A} state is stored in the odd rows of the detector while the light observed in the \textit{B} state 
(where the measured polarization has changed sign) is stored in the even detector rows. 
For the next frame (the \textit{$\pi$}-phase), the storage order is reversed: \textit{A} is saved in the even rows, \textit{B} in the odd rows.
Therefore the 4 exposures per $\theta_\mathrm{hwp}$ consist of $2$ `\textit{0}-frames' and $2$ `\textit{$\pi$}-frames'.

\subsection{Polarimetric Di\ff erential Imaging (PDI)}
We applied di\ff erent f\mbox{}ilters to the two beams of the beam splitter.
Therefore, we perform the reduction for each beam separately,
with our custom-made reduction pipeline.
After dark subtraction and f\mbox{}lat f\mbox{}ield correction, odd and even pixel rows are extracted 
into FLC \textit{A} and \textit{B} frames,
which we subtract for the \textit{0}-frames ($A_0 - B_0$) and the \textit{$\pi$}-frames ($A_\pi - B_\pi$).
Next, we subtract the \textit{$\pi$} di\ff erence images from the \textit{0} ones, and stack the 2 resulting images for each subsequent $\theta_\mathrm{hwp}$.
For $\theta_\mathrm{hwp} = 0^\circ$ we obtain the intensity ($I_{Q^+}$) and linear polarization ($Q^{+}$) by
combining \textit{0} and \textit{$\pi$} images:
\begin{eqnarray}
	I_{Q^+} &=&  0.5 \times ((A_0 + B_0) + (B_{\pi} + A_{\pi}))\big|_{\theta_\mathrm{hwp} = 0},\\
	Q^+     &=&  0.5 \times ((A_0 - B_0) - (B_{\pi} - A_{\pi}))\big|_{\theta_\mathrm{hwp} = 0}.
\end{eqnarray}
Similarly, for $\theta_\mathrm{hwp} = 45^\circ$ we obtain $I_{Q^-}$ and $Q^-$; $\theta_\mathrm{hwp} = 22.5^\circ$ gives $I_{U^+}$ and $U^+$;
and $\theta_\mathrm{hwp} = 67.5^\circ$ yields $I_{U^-}$ and $U^-$. 
To correct for charge trapping (a problem inherent to charge shu\ff \mbox{}ling, described by \citet{1994ApOpt..33.4254P} and \citet{2012SPIE.8446E..8YS}),
the afore mentioned procedures are done before any centering routine is applied.

The initial separation of the odd and even pixel rows leads to di\ff erent plate scales for the vertical and horizontal axes.
We obtain `square' $15 \times 15$\,mas pixels by binning $4 \times 2$ pixels.
We center the images by cross correlating the $I_{\theta_\mathrm{hwp}}$ images with a centered Mo\ff at function and apply
the same shift to the corresponding $Q^{+/-}$ or $U^{+/-}$ image.
The f\mbox{}inal Stokes components are:
\begin{eqnarray}
	I_Q &=&  0.5 \times (I_{Q^+} + I_{Q^-}),\\
	Q &=& 0.5 \times (Q^+ - Q^-),  
		\label{eqn:qtdif}\\
	I_U &=&  0.5 \times (I_{U^+} + I_{U^-}),\\
	U &=& 0.5 \times (U^+ - U^-),
		\label{eqn:utdif} 
\end{eqnarray}
from which we can compute the  Polarized Intensity $PI = (Q^2 + U^2)^{1/2}$, 
and the polarization angle $P_\theta = \arctan{(U/Q)}$. 

\subsubsection{Correcting Instrumental Polarization}
\label{sec:ip}
Computing the di\ff erence $Q^+  - Q^- $ images (same for $U^{+/-}$) with Equations~\ref{eqn:qtdif} \& \ref{eqn:utdif}
corrects for Instrumental Polarization (IP) created downstream from the HWP 
\citep[][C11]{Witzel11, 2014SPIE.9147E..87D, 2011A&A...531A.102C}.
However, this does not remove IP induced by the third mirror (M3) of the telescope and the f\mbox{}irst mirror (M4) of SPHERE, both of which are upstream from the HWP.
We cannot distinguish between this instrumental and real (inter)stellar and/or disk polarization at the location of the star.
C11 describe the correction for IP in imaging polarimetry, which assumes that the central star is unpolarized. 
Therefore we consider any signal measured over a small aperture at the center to be IP, which is the best we can do. 
Figure~\ref{fig:ipc} shows the IP corrected 
$PI$ images for $R'$-band (\textit{left}), $I'$-band (\textit{center}) and $R' + I'$ combined (\textit{right}).
The purple lines in the $R'$ and $I'$-band images show the direction of $P_\theta$.

\begin{figure}
   \begin{center}
   \includegraphics[width=0.5\textwidth,trim = 15 45 0 38]{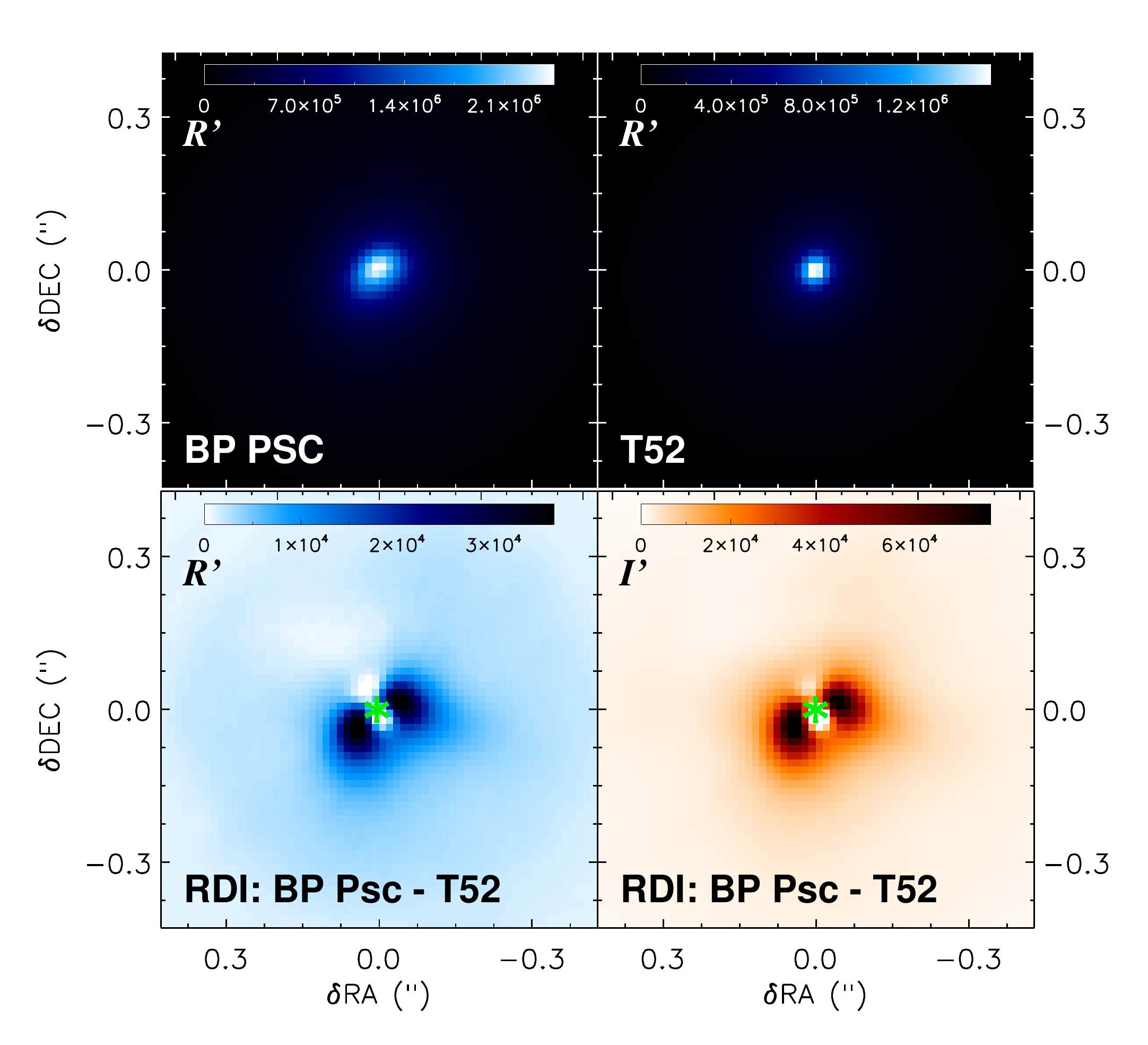}
   \end{center}
   \caption{ \label{fig:rdi}
   \textit{Top:} $R'$ band $I_t$ image of \bpp (\textit{left}) and T52 (\textit{right}). 
   \textit{Bottom:} RDI ($I_{t,\mathrm{\bppb}} - I_{t,\mathrm{T52}}$) images of the disk in $R'$-band (\textit{left}) and $I'$-band (\textit{right}).  
   The green asterisk shows the position of the star.
   } 
\end{figure} 
\subsection{Reference Star Di\ff erential Imaging (RDI)}
\label{sec:rdi}
When we compare the Stokes I (or total intensity $I_t =0.5 \times (I_Q + I_U)$)
images of \bpp and the point source T52 in the two top panels of Figure~\ref{fig:rdi}, 
we already see that \bpp is not just a point source: the disk contributes a signif\mbox{}icant part of the signal.
T52 was observed at lower air mass than \bpp (see Section~\ref{sec:obs}), 
which results in a smaller FWHM and higher Strehl ratio for T52. 
For Reference star Di\ff erential Imaging \citep[RDI][]{Smith:1984Sci}, where we subtract a reference star from our target,
this di\ff erence in FWHMs becomes a problem, which could lead to over-subtraction (i.e. removal of disk light) close to the star 
and under-subtraction (insu\ff \mbox{}icient removal of stellar speckle halo) at larger separations.

To adjust the Point Spread Function (PSF) of T52 to the lower Strehl ratio of the \bpp observations, 
we convolve T52 with a Gaussian of FWHM $=15$\,mas
to match the width of the $I_t$ image of \bpp at \textit{PA}\,$ = 30^\circ$, which is roughly 
perpendicular to the disk \textit{PA} of Z08. 
In this direction, we expect the inf\mbox{}luence of the disk on the shape of the PSF to be negligible.
Finally, we scale the peak f\mbox{}lux of the reference PSF to match the peak value for \bpp and 
subtract the former from the latter.
The bottom panels of Figure~\ref{fig:rdi} show the RDI images for $R'$ (left) and $I'$ (right).

\subsection{Deconvolution of the total intensity image}

\begin{figure}
   \begin{center}
   \includegraphics[width=0.48\textwidth,trim = 20 35 -15 45]{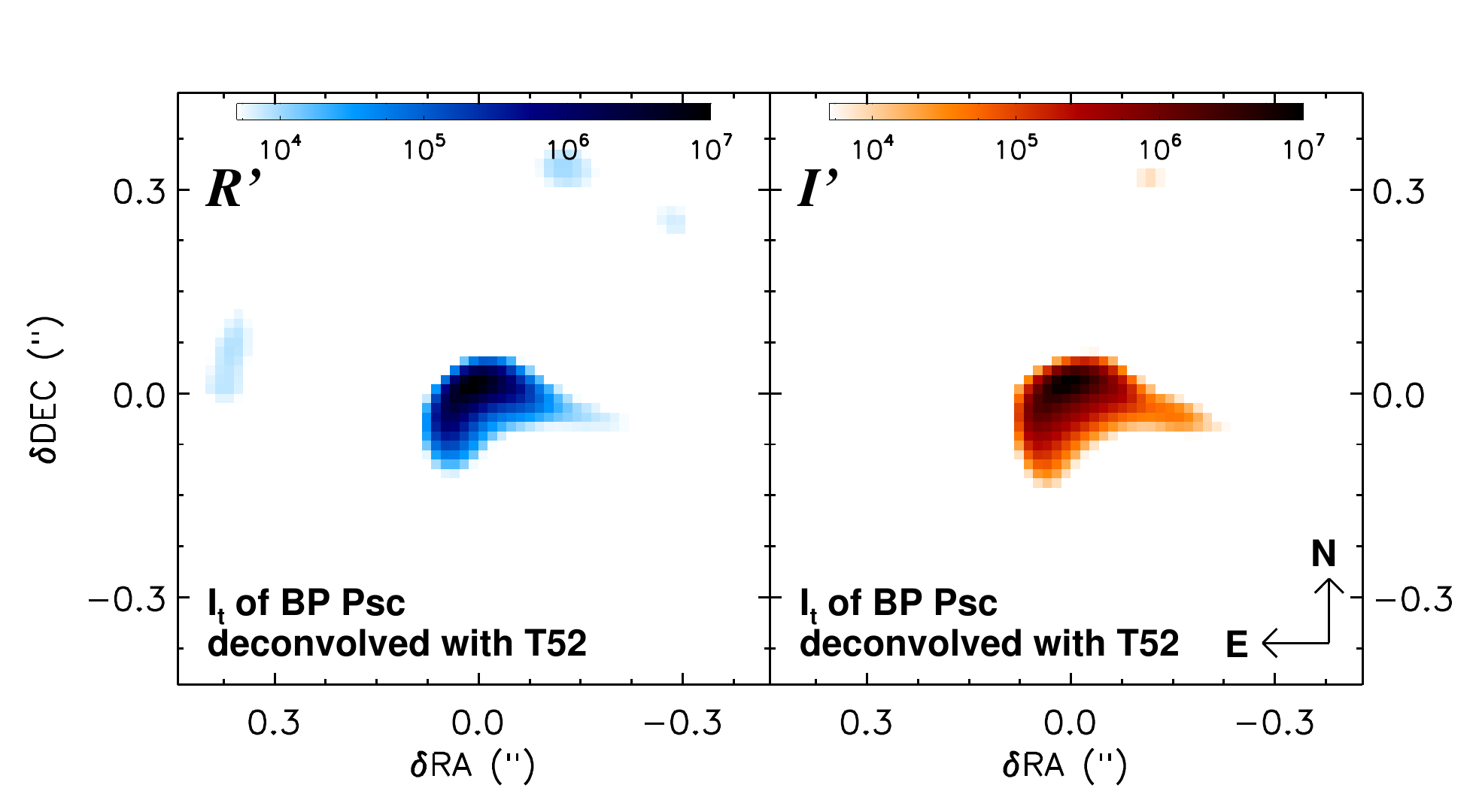}
   \end{center}
   \caption{ \label{fig:deconv}
   	$I_t$ image of \bpp deconvolved with the $I_t$ image of T52 in $R'$-band
(\textit{Left}) and $I'$-band (\textit{Right}). The deconvolution was performed with the Richardson-Lucy method.
   } 
\end{figure}

The only images of the resolved disk known to date are the deconvolved images of Z08, 
which display a di\ff erent structure than our $PI$ images in Figure~\ref{fig:ipc}.
To conf\mbox{}irm the detection of Z08, we apply the same method and 
perform the Richardson-Lucy (RL)
deconvolution \citep[using Equations~19 and 20 in][]{Lucy:1974AJ}
of the $I_t$ image of \bppb,
for which we show in Section~\ref{sec:rdi} that the disk signal forms a substantial part.
The observed $I_t$ is used as a starting guess for the deconvolution.
As in Section~\ref{sec:rdi}, the PSF is obtained by convolving the $I_t$ image of T52 with a Gaussian 
(FWHM $= 15$\,mas). 
RL recovers structures on scales larger than the FWHM within only a few ($\sim 100$) iterations.
Because our FWHM is comparable to the angular size of the disk we require more iterations to converge.
To monitor the convergence of the deconvolution 
we convolve the deconvolved images and subtract this from the original $I_t$ images.
After 2000 iterations, the residuals show little change. 
Figure~\ref{fig:deconv} shows the f\mbox{}inal RL deconvolved images for both f\mbox{}ilters.

\section{Results}
\label{sec:res}
We detect the disk of \bpp for the f\mbox{}irst time in $PI$ in both $R'$ and $I'$-bands (Figure~\ref{fig:ipc}).
Additionally, we retrieve the intensity image of the disk with RDI in Figure~\ref{fig:rdi}, 
and conf\mbox{}irm the detection of Z08 with the very similar results of our deconvolution (Figure~\ref{fig:deconv}).
The overall appearance of the $PI$ images of Figure~\ref{fig:ipc} is the same for both f\mbox{}ilters, 
and resembles a (nearly) edge-on f\mbox{}laring disk, similar to e.g. HH\,30 \citep[CTTS with $i  > 80^\circ$]{Burrows:1996ApJ}.
To give a f\mbox{}lux calibrated surface brightness of the disk requires a detailed calibration of
(e.g.~the polarimetric e\ff \mbox{}iciency of) ZIMPOL, which is beyond the scope of this paper. 
We conf\mbox{}ine this study to a qualitative description and analysis of the observations.
The main features are highlighted in the right-hand panel of Figure~\ref{fig:ipc}. 
A bright blob appears in the center of the images (feature \textbf{a}) covering the position of the star
(highlighted with a white asterisk).
The two bright regions (at position angles (\textit{PA}\,$ \sim 160^\circ$ and $ \sim 270^\circ$) within 0.15'' from the star (feature \textbf{b}) 
show the `bowl' of a f\mbox{}lared disk south of the mid-plane. 
This southern bowl dominates both the RDI images (Figure~\ref{fig:rdi}) and the deconvolutions (Figure~\ref{fig:deconv}) as well.
For the northern counterpart of this bowl, only two elongated regions are detected in the $PI$ images (features \textbf{c} in Figure~\ref{fig:ipc}).
However, two faint extended `f\mbox{}ingers' (feature \textbf{d} in Figure~\ref{fig:ipc}), starting at the features c, going outward to $\sim 0.5''$ seem to trace the extended surface of the f\mbox{}lared disk, north of the mid-plane.
Less clearly, similar f\mbox{}ingers can be seen extending outward to the south and west from the b regions.

Compared to high Strehl ratios reached for much brighter, $R < 8$\,mag
stars \citep[$S_{R'} \approx 60\%$,][]{Fusco:2014eo}, where high angular resolutions are reached 
\citep[e.g. FWHM$_{R'} \lesssim 20$\,mas for][]{Kervella:2015A&A},
the low Strehl ratios reached for \bpp ($R \approx 11$\,mag, $S_{R'} \approx 2.8\%$) 
naturally result in lower resolution: FWHM$_{R'} = 50$\,mas.
Both RDI and RL deconvolution are very sensitive to errors in the PSF. 
Due to the varying Strehl ratios, T52 can only be considered as an approximate PSF
(e.g. we have to broaden T52 to f\mbox{}it the FWHM of \bppb).
Therefore, we do not put too much emphasis on either the deconvolutions or the RDI images, 
we only emphasize that our deconvolved images in Figure~\ref{fig:deconv} look very similar to the ones obtained by Z08 in $H$ and $K$ band.
We base the main focus of our analysis on the $PI$ images in Figure~\ref{fig:ipc}.

\section{Discussion}
\label{sec:dis}
\subsection{Modeling the disk}
\label{sec:mod}

To f\mbox{}ind the evolutionary stage of \bppb, ideally we would compare disk models to those typical for disks around CTTSs 
and those around G-giants. 
To the best of our knowledge no radiative transfer models exist for disks around other f\mbox{}irst-ascent giants.
Models exist for more evolved systems
\citep[e.g.][]{Jeffers:2014A&A, Kervella:2015A&A}
However, a disk around a f\mbox{}irst-ascent giant must have formed very recently, 
which makes our knowledge on the large scale geometry of such systems very uncertain.
Therefore, we do not consider a comparison of this system to AGB disk models as a viable option.
Instead, we choose to treat \bpp as if it were a CTTS, at $d = 81$\,pc,
represented by a Kurucz stellar model atmosphere \citep{Kurucz:1979}
with surface temperature $T = 5000$\,K, luminosity $L = 0.67$\,$L_\odot$, and mass $M = 1.1$\,$M_\odot$,
and see if our best model resembles those created for disks around other CTTSs.
The disk radii and scale height ($H$) do essentially scale with distance, while leaving the overall shape unchanged. 
We do not aim to 
describe the dust properties of the disk, rather constrain the disk geometry 
by creating a model that f\mbox{}its the SED, while producing images that capture the major features of our observations. 

We use the Monte Carlo radiative transfer code MCMax 
\citep{Min:2009}\footnote{
see the MCMax website:
\href{http://www.hetisikke.nl/mcmax/}{http://www.hetisikke.nl/mcmax/}, 
or the manual: \href{https://sites.google.com/site/manualmcmax/home}{https://sites.google.com/site/manualmcmax/home}.
}
to f\mbox{}it the SED of \bppb. 
The disk models are made up of silicates and contain carbon with a carbon mass fraction $M_\mathrm{carbon}/(M_\mathrm{silicates} + M_\mathrm{carbon}) = 0.1$,
which is comparable to the fraction in the solar system.
Grain sizes lie between $0.05\,\mu$m and 3\,mm and decrease with a power law.
Similar to the results of M10, 
our models contained two disk components:
An inner disk at $0.12$\,au $\le r \le 1$\,au
and an outer disk at $r \ge 1.5$\,au. 
The scale height ($H$) of both ($j =$\, inner; outer) disk components increases with a power law $H(r) \propto r^p_j$ (Equation~18 of M09). 
The inner disk maintains this power law component ($p_\mathrm{in}$) for all radii. 
The increase of the outer disk goes with the power of ($p_\mathrm{out}$) up to a radius $r_\mathrm{exp}$ 
from where $H$ declines exponentially, causing a rapid decrease in the scattering of starlight.
The scale height decreases with particle size, due to settling of large dust grains towards the disk mid-plane.
Dust settling is higher (or $H_\mathrm{dust}/H_\mathrm{gas}$ is smaller) when turbulent mixing decreases.
Turbulent mixing is described by the viscosity parameter 
$\alpha_\mathrm{turb}\propto (H_\mathrm{dust}/H_\mathrm{gas})^2$ \citep{Dubrulle:1995Icar, Woitke:2016A&A}. 

We allowed the  inclination ($i$) to vary such that the line of sight optical depth at $550$\,nm $\tau_{550} = 5.2$, 
and found an inverse relationship between $H$ and $i$.
Even though we created f\mbox{}its to the SED with similar quality with a variety of parameters, 
the ZIMPOL observations placed lower limits on $H$, and therefore on $p_\mathrm{out}$, 
which in turn required an $i \lesssim 80^\circ$.
A degeneracy exists between the disk dust mass ($M_\mathrm{dust}$) and $\alpha_\mathrm{turb}$ for the f\mbox{}it of the Near and Mid IR range, where lower $M_\mathrm{dust}$ require higher values of $\alpha_\mathrm{turb}$ (i.e. settling becomes less e\ff \mbox{}icient).
The mass of the outer dust disk has a lower limit beneath which $i$ is no longer a\ff ected by $p_\mathrm{out}$ and only near edge-on conf\mbox{}igurations could still maintain the $\tau_{550}$ requirement.
This lower limit of $M_\mathrm{dust, out} \gtrsim 
10^{-7}$\,$M_{\odot}$ e\ff ectively gives us an upper limit for the viscosity parameter $\alpha_\mathrm{turb} \lesssim 5 \times 10^{-5}$. 
This means that the dust settling in the disk of \bpp is much stronger than the example of strong settling given by
\citet{Woitke:2016A&A}, who use $\alpha_\mathrm{turb} = 10^{-4}$.

\begin{figure}
   \begin{center}
   \includegraphics[width=0.5\textwidth ,trim =23 5 -20 20]{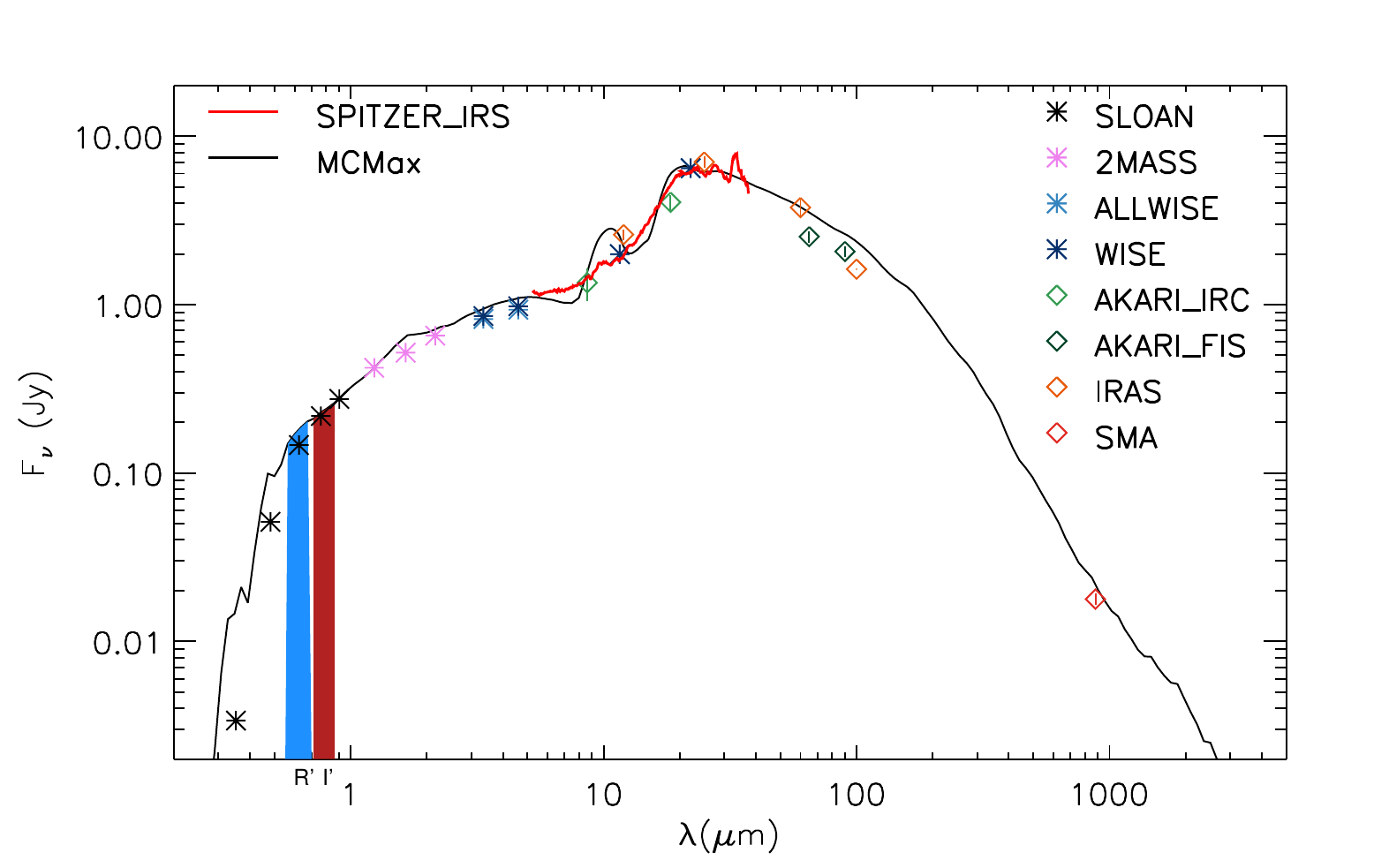}
   \includegraphics[width=0.5\textwidth,trim = 30 45 -20 0]{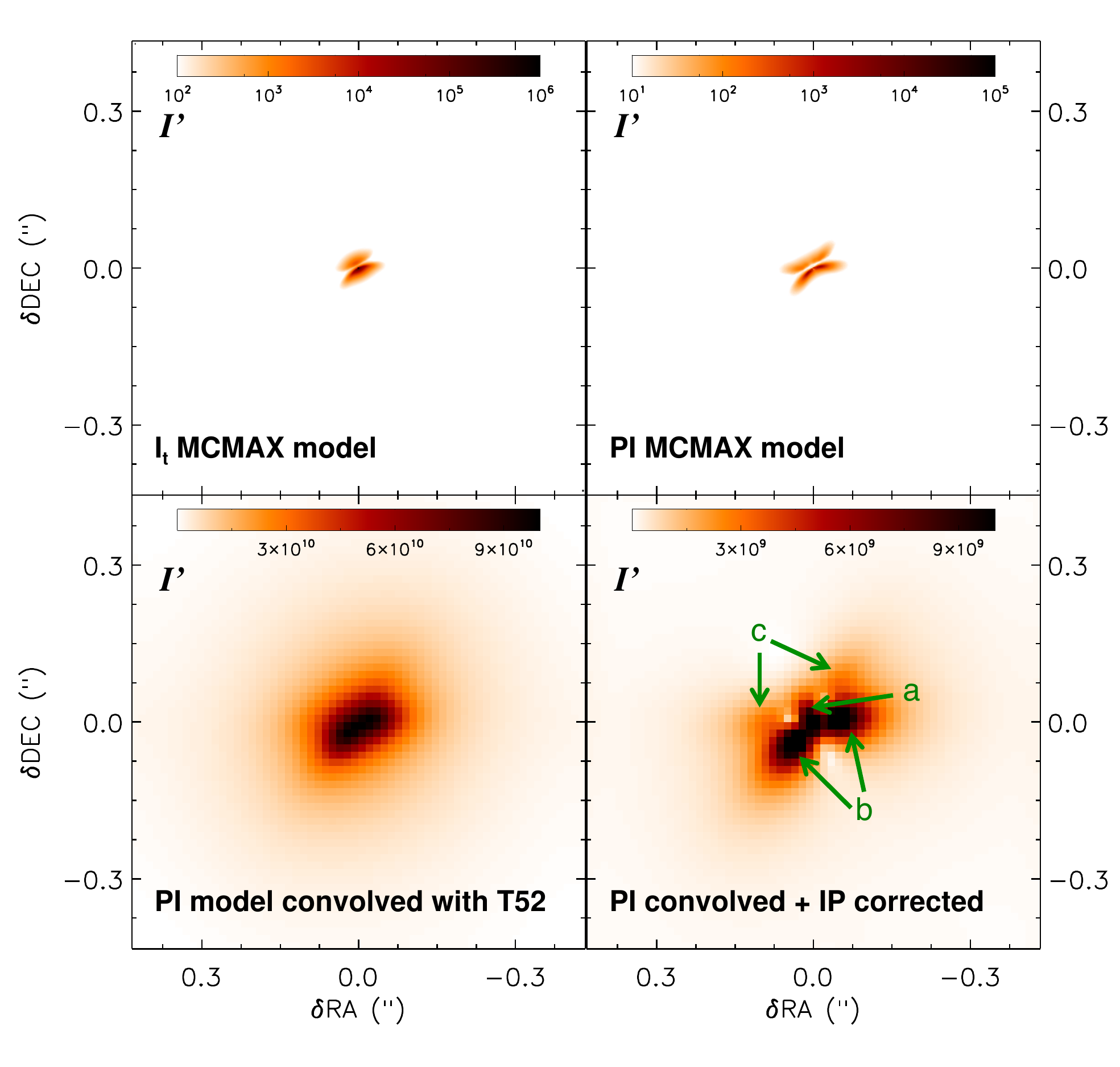}
   \end{center}
   \caption{ \label{fig:model}
   	\textit{Top}:
   	Spectral Energy Distribution of the MCMax model along the line of sight at $i = 78.9^\circ$ 
  compared to the photometric measurements and IRS spectrum.
  The blue and red bars indicate the ZIMPOL $R'$ and $I'$ f\mbox{}ilter bands respectively.
	The center and bottom row show the images corresponding to the MCMax Model. \textit{Center} in log scale: Not accounted for telescope resolution and IP the $I_t$ image (\textit{left}),
  	and the $PI$ image (\textit{right}). 
	\textit{Bottom} in linear scale: The same $PI$ model image after convolution with the reference star (\textit{left}); 
	the image on the \textit{right} is subsequently corrected for the signal at the center, (as if it were IP) as described in Section~\ref{sec:ip}. 
	The features a, b, and c are comparable to the features with the same labels in 
	Figure~\ref{fig:ipc}.
   } 
\end{figure} 
The f\mbox{}inal model (Figure~\ref{fig:model})
has $i = 78.9^\circ$ (in agreement with $i = 75 \pm10^\circ$ of Z08); 
$r_\mathrm{exp} = 30$\,au; $\alpha_\mathrm{turb} = 5.0 \times 10^{-6}$; and $p_\mathrm{in} = 1.14$ and $p_\mathrm{out} = 3.0$ for the inner and outer disk respectively.
The model f\mbox{}
its the SED very well, except for a strong silicate feature at $10$\,$\mu$m.
We have found that increasing the carbon/silicate fraction can remove this feature, 
but 
using this chemical composition
we did not achieve a good f\mbox{}it to the SED overall so far.
Since we consider the chemical composition of the disk to be outside of the scope of this paper, 
we did not pursue the removal of the silicate feature any further.
The Stokes images of the disk model in the direction of the line of sight are convolved with the $I_t$ image of the reference star,
after which we applied the same procedure of IP correction as we did for the observations. 
The model will not truly be plagued by such an artifact, 
but as we explain in Section~\ref{sec:ip}, we cannot distinguish between true polarization at the location of the star and IP in our observations. 
Applying the IP correction therefore ensures the best comparison between model and observation 
\citep[as demonstrated by][]{Min:2012A&A}.

Figure~\ref{fig:model} shows for $I'$-band the unconvolved model $I_t$ and $PI$ images (top left and right respectively), 
and the $PI$ image after convolution with the PSF (bottom left), and IP correction (bottom right).
The convolved + IP corrected image shows a striking similarity with the reduced $PI$ images of Figure~\ref{fig:ipc}.
The features a, b and c, described in Section~\ref{sec:res}, are visible in the convolved + IP corrected model image.
Comparing the top-right and bottom-right images of Figure~\ref{fig:model} teaches us that features b and c represent the upper and lower arcs of a nearly edge on disk, while feature a is an artifact, caused by the convolution + IP correction.
We did not achieve to create a model with features, similar to d. 
However, from the morphology of the unresolved disk model, 
we can deduce that they are likely to be extensions of the northern arc of the disk.
This explanation is supported by the fact that the c features lie at the base of the d features.
In an alternative explanation the second disk component is surrounded
by a third and outermost disk component.

The angle between the northern and southern arc ($\sim 2H/r$) seems to be smaller for the model than for the ZIMPOL observations.
Therefore, we do not claim to have found a unique solution for either $i$ or $H$ of the disk.
We rather created a model with a morphology comparable enough to explain the observations.
Our model is strongly f\mbox{}lared for its small grains, but extremely f\mbox{}lat for the larger grains.
ALMA long baseline observations should be able to conf\mbox{}irm if our modeling results are correct. 
Since small dust grains are coupled to the gas, 
we predict that the molecular disk shows a much stronger f\mbox{}laring than the large grain dust disk.

The parameters (especially f\mbox{}laring and settling) used for the presented model are far from the classical Keplerian disk in hydrostatical  equilibrium,
which means that the disk of \bpp is not a stereo-typical protoplanetary disk for a CTTS. 
This could be explained by a very young star, with infall from its native star-forming nebula. 
However, since no associated star-forming region has been found, we do not consider this to be a plausible scenario.
An alternative explanation for the atypical behavior of the disk is that it is not a protoplanetary disk surrounding a pre-main sequence star,
but rather a disk around a f\mbox{}irst-ascent G-giant.
Unfortunately we lack G-giant disk models which allow for a proper comparison.
An open question remaining for the G-Giant scenario is whether the formation of the disk in a way as proposed by
Z08 and M10 (i.e. enveloping a massive companion) is compatible with the strong f\mbox{}laring of small grains and settling of the larger grains. 

\subsection{Disk morphology}
\label{sec:morph}

Even though both sides of a strongly inclined (but $i \ne 90^\circ$) disk will be dominated by forward scattering,
we expect the forward facing side (or top) to be brighter than the backward facing side (or bottom).
The asymmetry originates from the smaller optical depth of the forward facing side of the disk (the starlight reaches us more e\ff \mbox{}iciently).
From the MCMax model images, we determine that the top side of the disk is pointing south and the bottom side north.
An interesting test to conf\mbox{}irm that the southern side is facing us will be to check 
whether the associated southern HH-object (Z08) is blue shifted, and the northern counterpart red shifted.

We determine the \textit{PA} by assuming (as we see for the disk model images in Figure~\ref{fig:model}) 
that the largest symmetry will be acros the axis \textit{PA}\,$\pm\,90$, while the symmetry acros the \textit{PA} will always be broken for a
disk with $i \ne 90^\circ$.
We mirror the disk image along the \textit{PA} and subtract the mirror image from the original disk image.
For regions where the SNR is high in the original image, we take the absolute value of the residual (image - mirror).
We repeat this method for varying \textit{PA}.
The angle which provides the smallest residual signal yields \textit{PA}\,$= 120.8 \pm 2.0^\circ$, 
in good agreement with Z08 (\textit{PA}\,$= 118 \pm 5^\circ$).

\section{Conclusion}
\label{sec:con}

Our ZIMPOL observations of \bpp conf\mbox{}irm the presence of a circumstellar disk.
Despite a modest AO correction, we resolved the disk for the f\mbox{}irst time in the visible ($R'$ and $I'$ band), 
and present the f\mbox{}irst polarimetric images of this object. 
Our deconvolved image in Figure~\ref{fig:deconv} conf\mbox{}irms the disk images of Z08,
and retrieve the \textit{PA} with a higher accuracy than was known until now.

The MCMax modeling yields images comparable to the observations. 
They require a model which is strongly f\mbox{}lared for small grains, yet strong settling occurs for large grains.
Both f\mbox{}laring and settling values are atypical for a proto-planetary disk of a T-Tauri type star. 
Without strongly discarding the CTTS scenario, our study therefore is more inclined towards a G-giant evolutionary stage for this system.
The comparison between model and observations allows us to determine that the forward facing side is pointing south 
($\approx 210.8^\circ$), while the backward facing side points north ($\approx 30.8^\circ$).
The specif\mbox{}ic disk features detected in Figure~\ref{fig:ipc} can be explained by the model as either resolved components of the 
forward facing side of the disk (`b' features); the backward facing side (`c' \& 'd' features) or as a residual of the convolution with the telescope PSF and IP subtraction (`a' feature).

\section*{Acknowledgements}
We are thankful to the ESO support sta\ff~on Paranal and the SPHERE science verif\mbox{}ication team for their support and the successful observations.
H.C. acknowledges support from the Spanish Ministerio de Econom\'ia y Competitividad under grant AYA2014-55840P.




\bibliographystyle{mnras}
\bibliography{ref161007} 



%
%
%


\bsp	
\label{lastpage}
\end{document}